# NdAlSi: a magnetic Weyl semimetal candidate with rich magnetic phases and atypical transport properties

Jin-Feng Wang[1,2], Qing-Xin Dong[2,3], Zhao-Peng Guo[2,3], Meng Lv[2,3], Yi-Fei Huang[2,3], Jun-Sen Xiang[2,3], Zhi-An Ren[2,3,4], Zhi-Jun Wang[2,3,4], Pei-Jie Sun[2,3,4]*, Gang Li[2,3,4]* and Gen-Fu Chen[2,3,4]*

[1] *Henan Normal University, College of Physics, Xinxiang, Henan 453007, China*

[2] *Institute of Physics and Beijing National Laboratory for Condensed Matter Physics, Chinese Academy of Sciences, Beijing 100190, China*

[3] *School of Physical Sciences, University of Chinese Academy of Sciences, Beijing 100049, China*

[4] *Songshan Lake Materials Laboratory, Dongguan, Guangdong 523808, China*

*To whom correspondences should be addressed. E-mails: pjsun@iphy.ac.cn
                                                     gli@iphy.ac.cn
                                                     gfchen@iphy.ac.cn

Magnetic Weyl semimetals (MWSM) have attracted significant attention due to their intriguing physical properties and potential applications in spin-electronic devices. Here we report the characterization of NdAlSi including transport, magnetization, and heat capacity on single crystals, as well as band structure calculation. It is a newly proposed MWSM candidate which breaks both time-reversal and spacial inversion symmetries. A temperature-magnetic field phase diagram is experimentally established. Remarkably, on the angular magnetoresistance (AMR), a two-fold symmetric sharp peak instead of a smooth variation is observed across the phase boundary between ferrimagnetic and antiferromagnetic phase. We argue that the tunability of both the topological and magnetic properties in NdAlSi is crucial for realizing such a behavior. Our results indicate that 4f-electron-based MWSM can provide a unique platform to explore new and intriguing quantum phenomena arising from the interaction between magnetism and topology.

## I. INTRODUCTION

Weyl semimetals (WSM) are solid crystalline materials hosting emergent fermionic quasiparticles known as Weyl fermions close to the nodes of two nondegenerate bulk bands. Such nodes are topologically protected, requiring broken of either spacial inversion or time-reversal symmetry (via magnetic order or an applied magnetic field) [1-6]. In the magnetic Weyl semimetals (MWSM), the magnetization can easily modify the Weyl nodes to induce Berry curvature field, resulting in novel phenomena such as intrinsic anomalous Hall effect, anomalous Nernst effect, and giant magneto-optical responses, as evidenced in $Co_3Sn_2S_2$, $Co_2MnGa$, $Mn_3Sn$, $EuCd_2As_2$ and $Fe_3GeTe_2$ [7-14]. These characteristics make MWSM a promising platform for potential application in spintronics and optical/electronic devices, and stimulate researchers to explore new systems with novel magnetic Weyl physics.

MWSM having a simultaneous breaking of both spacial inversion and time-reversal symmetries are relatively rare. Very Recently, the *R*Al*X* (*R*: Ce, Pr, Nd; *X*: Si or Ge) family comes into focus of research, which has a non-centrosymmetric crystal structure while the rare earth element *R* usually carries intrinsic f electron magnetic moment [15-16]. By rare earth element substitution, the family exhibits a great material tunability, nonmagnetic WSM (LaAlGe), type I (PrAlGe) and type II (CeAlGe) MWSM have been realized [15, 17-18]. Compared with other MWSM [19-20], Weyl nodes in *R*Al*X* are more robust and less dependent on the details of magnetism [15, 21], which acts as an effective Zeeman field that shifts the Weyl nodes in momentum space. Various ground states such as magnetic glassy phase, ferromagnetic (FM), ferrimagnetic (FIM) and antiferromagnetic(AFM) states, have been reported for the *R*Al*X* family [16-17, 22-24], make them appropriate candidates for investigating the relation between magnetism and topology. To date, the Ge-based RAlGe have been relatively well investigated [15, 22, 25-28], but investigations on the Si-based RAlSi are rare [16, 29].

In this work, we synthesized high-quality single crystals of NdAlSi and carried out magneto-transport, magnetization, and heat capacity measurements. Several magnetically ordered regions at low temperature are defined with magnetic field as a tuning parameter, while band structure calculation confirms the existence of Weyl nodes among different magnetic structures. As a consequence of the tunability, a sharp peak is observed on the angular magnetoresistance (AMR) when external magnetic field is applied along a direction perpendicular to the easy axis of magnetization in the field-induced ferrimagnetic phase, which has not been widely observed and discussed in MWSM.

## II. EXPERIMENTAL AND CALCULATION DETAILS

Single crystals were grown by the Al self-flux method [30]. Raw materials of Nd, Al, and Si with a molar ratio of 1:10:1 were weighted, loaded into an alumina crucible, and then sealed in a quartz tube under high vacuum. The sealed quartz tube was heated to 1150 ℃, held for 6 hours, and slowly cooled to 750℃ at a ratio of 2℃/h. The excess of Al flux was removed by centrifuging. Finally, large shiny plates-like single crystals were obtained. A trace of residual Al on the surface of single crystals was removed by soaking in diluted NaOH solution. The crystal structure was examined by both powder and single-crystal x-ray diffractions. Resistivity and specific heat measurements were performed in a Quantum Design Physical Property Measurement System (PPMS). The magnetization was measured by a Quantum Design Magnetic Property Measurement System (MPMS).

The first-principles calculations were implemented in the Vienna *ab initio* simulation package (VASP) [31] using the projected augmented wave method. The Perdew-Burke-Ernzerh of generalized gradient approximation exchange-correlations functional was implemented in calculations [32]. The cutoff energy of the wave functions was 340 eV. The Brillouin zone was sampled by the Monkhorst-Pack method [33] with k-points spacing of $0.025 \times 2\pi$ Å$^{-1}$. The spin-orbit coupling (SOC) was taken into accout within the second varational method [34]. Considering the strong correlation effect in 4f electrons of Nd atoms, the Hubbard+U method [35] was implemented in the calculation. The maximally-localized Wannier functions was constructed using Wannier 90 package [36] and the positions of Weyl points were calcuated using Wannier Tools package [37]. The quantum oscillation frequencies were calculated using SKEAF program [38].

## III. RESULTS AND DISCUSSIONS

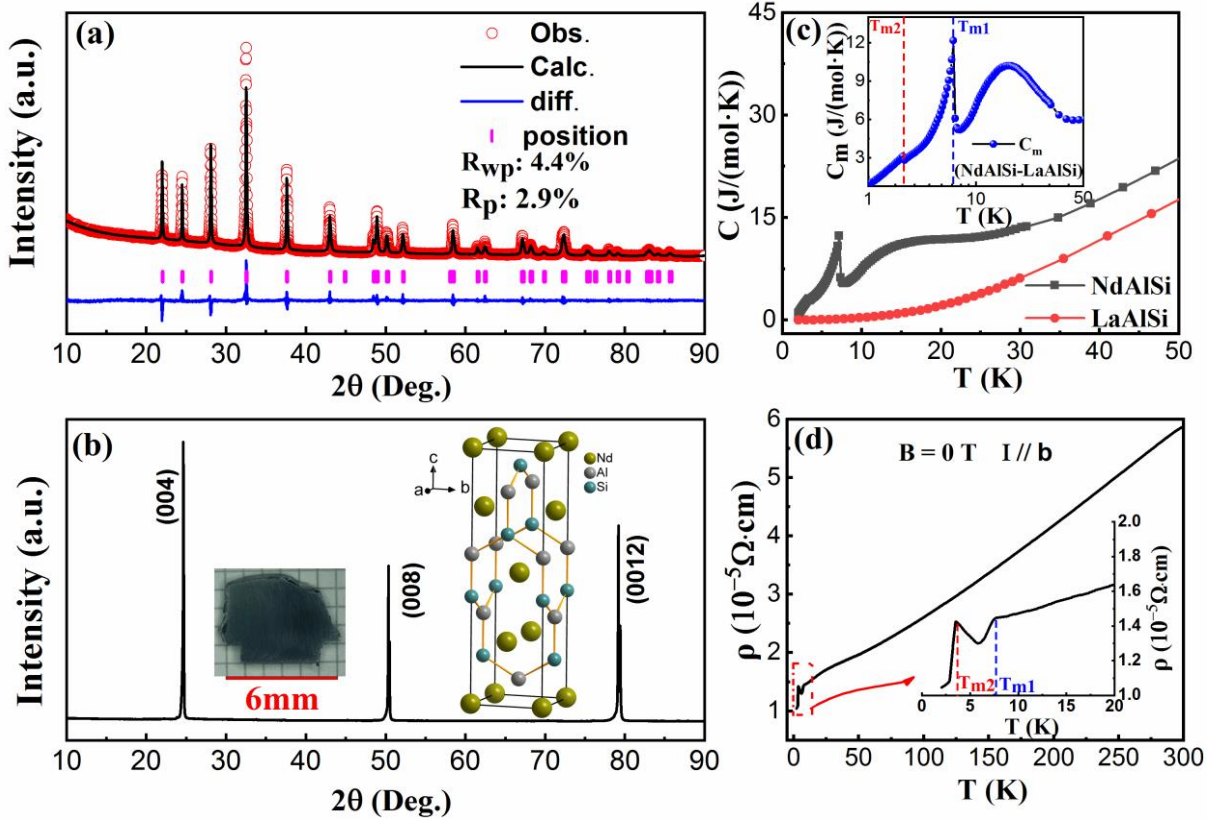

**FIG.** 1. Structure, specific heat and electrical resistivity characterizations of NdAlSi. (a) Powder XRD pattern measured at room temperature. Red circle and black solid line represent the collected (Obs.) and calculated (Calc.) data, blue line exhibits their difference (diff.), and the pink bars denote the positions of Bragg peaks, respectively. (b) Single crystal X-ray diffraction peaks, the insets show an optical image of a grown single crystal (left) and the LaPtSi-type ($I4_1md$) structure (right). (c) Specific heat of NdAlSi and LaAlSi from 2 K to 50 K, the inset gives the magnetic heat capacity $C_m$ (T) by subtracting the specific heat of LaAlSi from that of NdAlSi. (d) ρ (T) with the low temperature part enlarged as inset. In (c) and (d), those transition temperatures are marked by vertical dashed lines.

As exhibited in Fig. 1(a), all the diffraction peaks can be well indexed in a tetragonal LaPtSi-type crystal structure with non-symmetric space group $I4_1md$ (No.109), the refined lattice parameters are **a = b = 4.204(7) Å, c = 14.524(7) Å**, consistent with those reported in literature [39]. The best atomic ratio determined by Energy Dispersive X-ray spectroscopy (EDX) is Nd : Al: Si = 1 : 0.98 : 0.97, close to stoichiometry, which is consistent with the refinement results of single crystal (Table I in Appendix). Fig. 1(c) shows the temperature dependence of specific heat for the magnetic NdAlSi and nonmagnetic analogue LaAlSi at zero magnetic field. Their difference gives the magnetic specific heat $C_m(T)$ of NdAlSi as present in the inset of Fig. 1(c). With decreasing temperature, $C_m(T)$ exhibits a broad (Schottky type) peak centered at 18 K, contributed by the crystal electric field (CEF) splitting of $Nd^{3+}$ atomic levels, followed by a pronounced λ-shaped peak at 7.3 K (labeled as $T_{m1}$ hereafter) and a small kink-like feature at 3.5 K ($T_{m2}$). These two characteristic temperatures mark two magnetic phase transitions in NdAlSi, which are also evidenced in magnetic susceptibility $\chi(T)$ (Fig. 2(a)) and temperature dependent resistivity $\rho(T)$ (Fig. 1(d)). In $\chi(T)$, there exists a large magnetic anisotropy at low temperature between B//c-axis and B⊥c-axis, and $\chi_{//} / \chi_{\perp}$ goes up to ~45 at 2 K. $\chi_{//}(T)$ exhibits a subtle upturn at $T_{m1}$, followed by a significant jump up at $T_{m2}$, and tends to saturate at lower temperatures, while $\chi_{\perp}(T)$ shows two clear antiferromagnetic-like drops at both $T_{m1}$ and $T_{m2}$. For T >50 K, Curie-Weiss fits of $\chi^{-1}(T)$ for both B⊥c-axis and B//c-axis give an effective magnetic moment $\mu_{eff}$ ~3.7 $\mu_B$, close to the theoretically expected value for the free $Nd^{3+}$ ion (3.6 $\mu_B$), while the fitted Curie-Weiss temperatures $\theta_p$ for B⊥c and B//c are -13 K and 9.8 K (Appendix Fig. 7), respectively. In $\rho(T)$, the overall behavior is metallic, the resistivity shows a downward kink at $T_{m1}$, and then a sharp drop at $T_{m2}$ upon further cooling, without flat-out down to 2 K. However, $\rho(T)$ is not monotonic in-between $T_{m1}$ and $T_{m2}$, exhibiting a minimum around 6 K followed by an upturn as T towards $T_{m2}$. An increase of resistivity upon cooling through a magnetic transition is usually ascribed to either the change of the electronic band structure, i.e., a reduction of carrier density due to the formation of superzone gap or spin density wave (SDW), or the enhanced scattering inside the domain walls due to a misalignment of the spins [40-43]. It is plausible that this resistivity upturn is caused by the development of SDW gap at the Fermi surface due to emergence of an incommensurate antiferromagnetic (IC AFM) ordering [39].

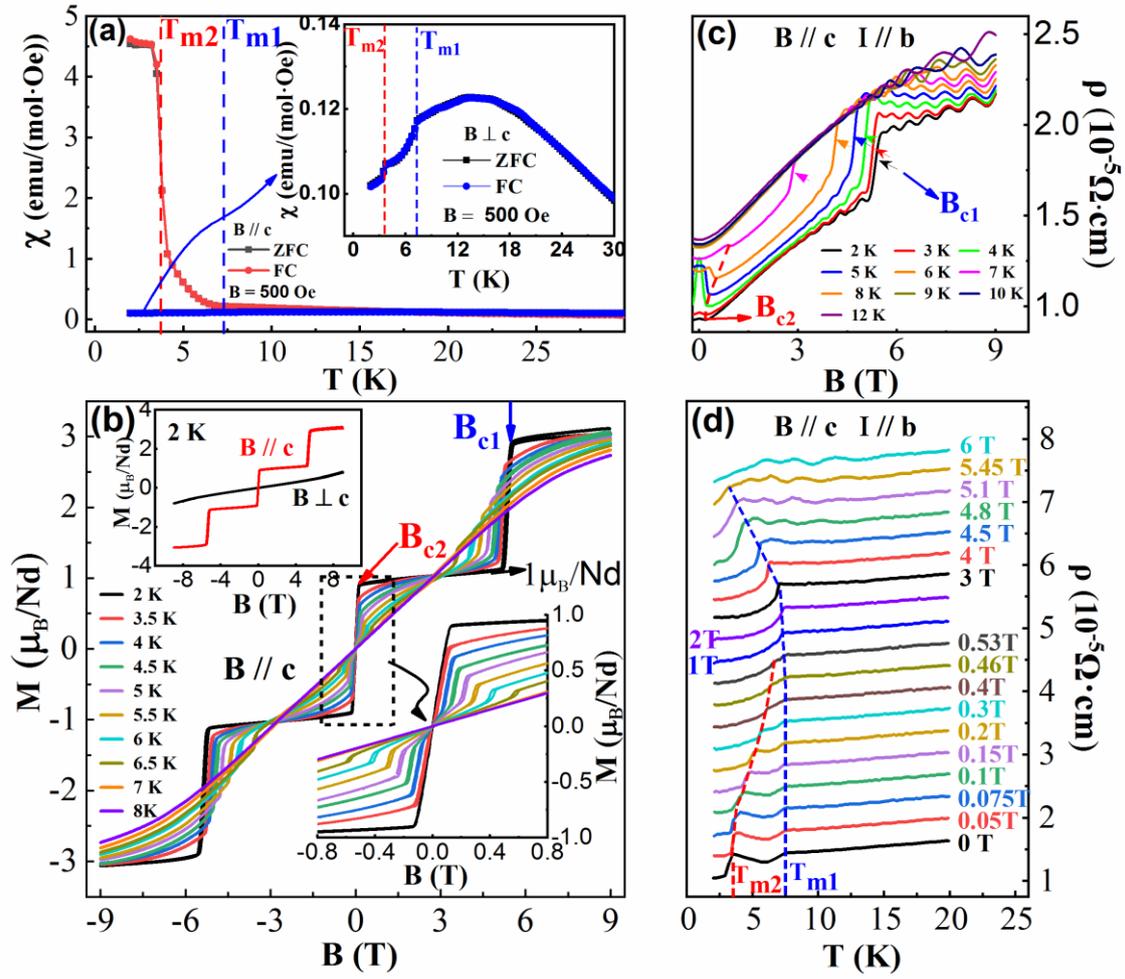

**FIG. 2.** Magnetic and transport properties of NdAlSi. (a) The dc magnetic susceptibility χ(T) measured at 500 Oe for B // c and B⊥c, respectively. (b) Isothermal M (B) curves for H//c-axis, an enlargement around zero field showing small hysteresis and a comparison with H⊥c-axis are included as insets. (c) Isothermal magneto-resistivity curves with B//c. (d) The evolution of ρ (T) at constant external magnetic fields for B//c. Data are vertically shifted by $0.35*10^{-5}$ Ω·cm for clarity.

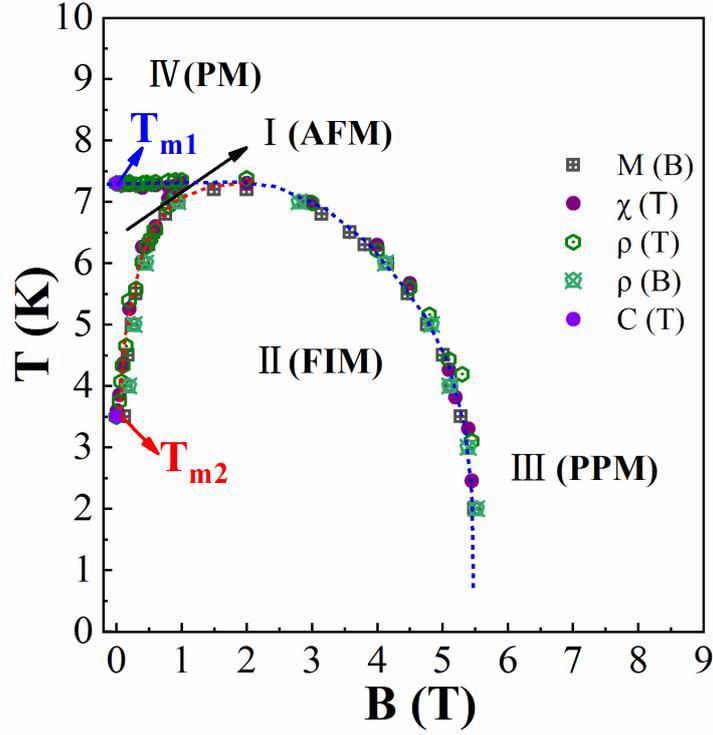

**FIG. 3.** The temperature-magnetic field phase diagram of NdAlSi for B//c axis. Denotations are in main text. Dotted lines are guide to eye.

For metallic NdAlSi, how the conducting charge carriers interact with the Nd 4f electrons is a key question in understanding its properties, and a crucial step is to establish its magnetic field-temperature phase diagram. Detailed field dependent magnetization and resistivity results are presented in Figs. 2(b)-2(d). As shown in the upper-left inset of Fig. 2(b), at 2 K the magnetization M for B⊥c is close to linear with field up to 9 T, while for B//c it has two step-like increases into plateaus, so the compound has a ferromagnetic-type order at low temperature with c-axis as the easy axis of magnetization. In the main panel of Fig. 2(b), the magnetic fields corresponding to the two steps for B//c are labeled as $B_{c1}$ (high B), and $B_{c2}$ (low B), respectively. Their evolution with temperature can be clearly traced. With increasing temperature, the size of the magnetization steps is suppressed, meanwhile, the two step-like transitions survive but gradually move close to each other, merge at T = 7 K then vanish. By normalizing magnetization to Bohr magneton, the low field plateau is about 1$\mu_B$/Nd$^{3+}$, while the high field plateau is about 3$\mu_B$/Nd$^{3+}$, which is very close to the full magnetization of free Nd$^{3+}$ ion (gJ =

3.27 $\mu_B$). The aforementioned first 1/3 plateau of the full magnetization observed can be reasonably deduced that originating from a ferrimagnetic state with an up-up-down (↑↑↓) spin alignment. By increasing magnetic field, a metamagnetic transition occurs at $B_{c1}$, which corresponds to a spin-flip transition for ferrimagnetically ordered spins [44-45], followed by the magnetization saturating to a ferromagnetic 3/3-plateau (↑↑↑). The picture is unusual for a square lattice system however has been confirmed by recent neutron diffraction result [39]. It is further shown in the inset in Appendix Fig. 8(a) that at 6 T the zero-field cooling (ZFC) and field-cooling (FC) M(T) curves overlap with each other, excluding the possibility of a ferromagnetic phase transition. Therefore, the magnetic state at low temperature for B > $B_{c1}$ is a field-induced polarized paramagnetic (PPM) region, which has been realized in those *f*-electron antiferromagnetic systems with spin density wave (SDW) [46].

The two characteristic fields $B_{c1}$ and $B_{c2}$ can also be traced on isothermal magnetoresistivity (MR) for B//c and I//b, as shown in Fig. 2(c). The close to zero field MR(B) has a conventional polynomial dependence on B at T > $T_{m1}$. While for $T_{m1}$ > T > $T_{m2}$, a negative MR(B) develops, gradually becomes pronounced with decreasing temperature, and reaches -28% at 0.2 T at 4 K, accompanying with a shift of $B_{c2}$ to lower fields. Such a resistivity behavior across critical field of $B_{c2}$ occurred at a moderate temperature can be ascribed to the gradual suppression of the possible energy gap at the Fermi surface due to the magnetic phase transition from the IC AFM to FIM state. Then for T < $T_{m2}$, a very small drop appears in resistivity at $B_{c2}$, which is believed to arise dominantly from the freezing out of spin-flip scattering due to the alignment of spins by the magnetic field [16]. At the same time, a sharp increase in resistivity exists at higher field for T < $T_{m1}$, the middle points of the upturn are identified as $B_{c1}$. Their evolution with temperature is consistent with that from isothermal M(B). Given that the full alignment of magnetic moments at B = $B_{c1}$ should suppress spin-flip scattering, considering this mechanism alone would lead to a reduction in resistivity when increasing field across the transition, which is opposite to the actual jump up at $B_{c1}$, thus other mechanisms, likely a Fermi surface reconstruction, is present. This argument is confirmed by analyzing the Shubnikov-de Haas (SdH) oscillations on the magnetoresistivity data. At 2 K, oscillations can be traced starting from ~2 T, in 2~5 T range the fast Fourier transform (FFT) on background-subtracted magnetoresistivity data yield two frequencies at 32 T and 66 T, respectively, while in 5.6-9 T range there exists another frequency ~86 T, marking a large change of the Fermi surfaces across the field induced transition at $B_{c1}$. Evolution of quantum oscillation frequencies with temperature and field for B//c axis is presented in Appendix Fig. 9. These cross-sectional areas of the Fermi surfaces extracted from SdH are all less than 0.1% of the projected area of the first Brillouin zone, indicating NdAlSi is a low carrier density system, whereas the resistivity at low temperature is on the order of 10 μΩ·cm, evidencing a high carrier mobility. As shown in Appendix Fig. 10, a single band fitting of Hall resistivity data within 0.5 T around zero yields a carrier density on the order of $10^{20}$ cm$^{-3}$, and mobility of $10^3$ cm$^2$/(V·s).

Figure. 3 gives the temperature-magnetic field phase diagram of NdAlSi for magnetic field along the c-axis, constructed based on the above heat capacity, magnetization, and resistivity data. At zero field, the compound enters a magnetically ordered region when cooled down below 7.3 K, the anisotropy of magnetization increases and the region is labeled as AFM in the sense that it is reminiscent of the incommensurate SDW found in a ferromagnet [46], in which the moments are organized ferromagnetically in local, however the amplitude is modulated and the phase is reversed among a distance larger than lattice constant. By cooling below 3.5 K, the moments are further aligned ferrimagnetically but with less entropy change. This picture is consistent with recent neutron result [39]. By turning on magnetic field, the phase line between the AFM and the ferrimagnetically ordered region (FIM) shifts, and further increasing field drives the compound from FIM to PPM. It is noted that on the current phase diagram whether there exists a boundary between the PM and the PPM region is not clear yet. In addition, in the PM-PPM region, a phase shifting of quantum oscillation is observed, as well as an oscillation of resistivity with temperature at fixed fields, as shown in Fig. 2(d), which would be further explored in our next studies.

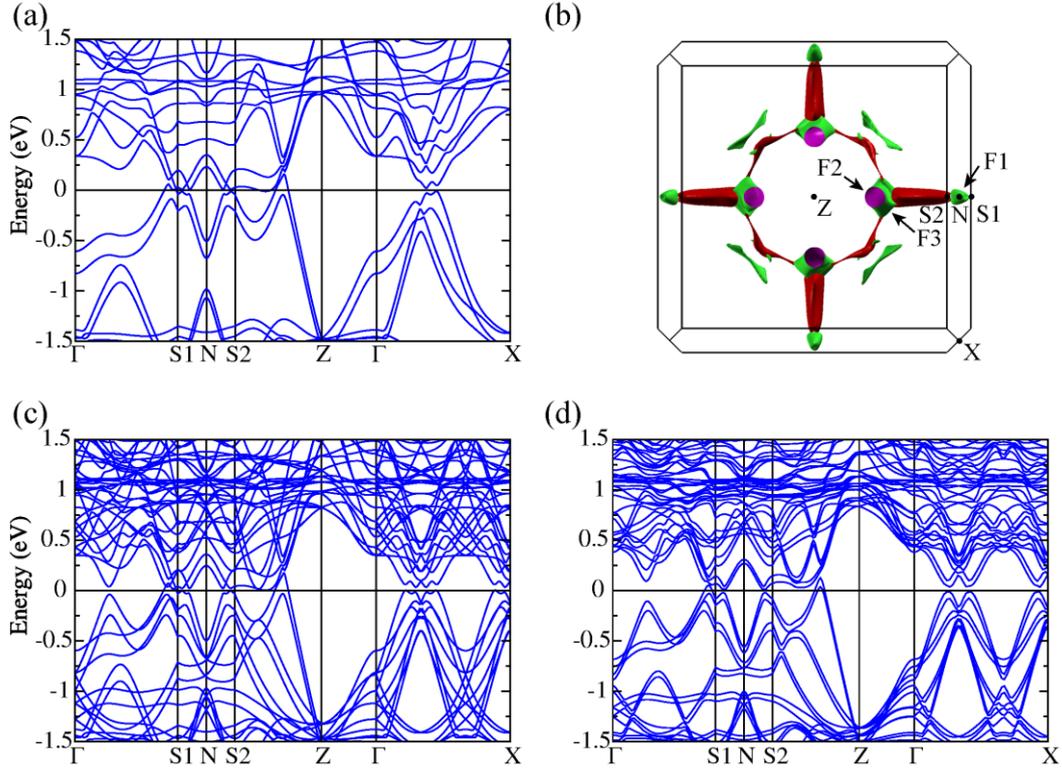

**FIG. 4.** The calculated band structure and Fermi surfaces of NdAlSi by DFT including spin orbit coupling (SOC). (a) The band structure of the PPM state, and (b) The corresponding Fermi surfaces. The locations of extreme orbits are marked by F1, F2 and F3. (c) The folded band structure of the PPM state in the FIM unit cell. (d) The band structure of the FIM state.

How the configuration of magnetic moments affects the band structure is tested by our calculations. We considered the PM, the PPM and the FIM states. The magnetic moment on Nd atoms are set along z direction with 3 $\mu_B$. Fig. 4(a) gives the band dispersion of the PPM state, in which there are 56 Weyl points, while in the PM state there are 40 Weyl points (Appendix Fig. 12). The three extreme cross-sections on the Fermi surfaces are labeled as F1, F2, and F3 in Fig. 4(b). The experimentally extracted quantum oscillation frequency, 86 T, is close to the frequency of F2, which is located in the $k_z = \pi$ plane and not enclose any Weyl point. However, after manually tuning the Fermi energy $E_F$ up by 20 meV, the frequency would be close to that of F3, locating at $k_z = 0.62\pi$ plane and close to a Weyl point, illustrating that Fermi surfaces of such a low carrier density system are quite sensitive to calculation details (Appendix Table II). However, the overall band structures of the PM, the PPM, and the FIM states do not change radically. As shown in Figs. 4(c) and 4(d), from the PPM to the FIM state, the band dispersions close to $E_F$ are similar, with the hole bands along S2-Z shrinks and the direct gap along Γ-S1 becomes larger. This is in accordance with literature that in this family of materials the magnet moments from f electrons serves as an effective Zeeman field [15].

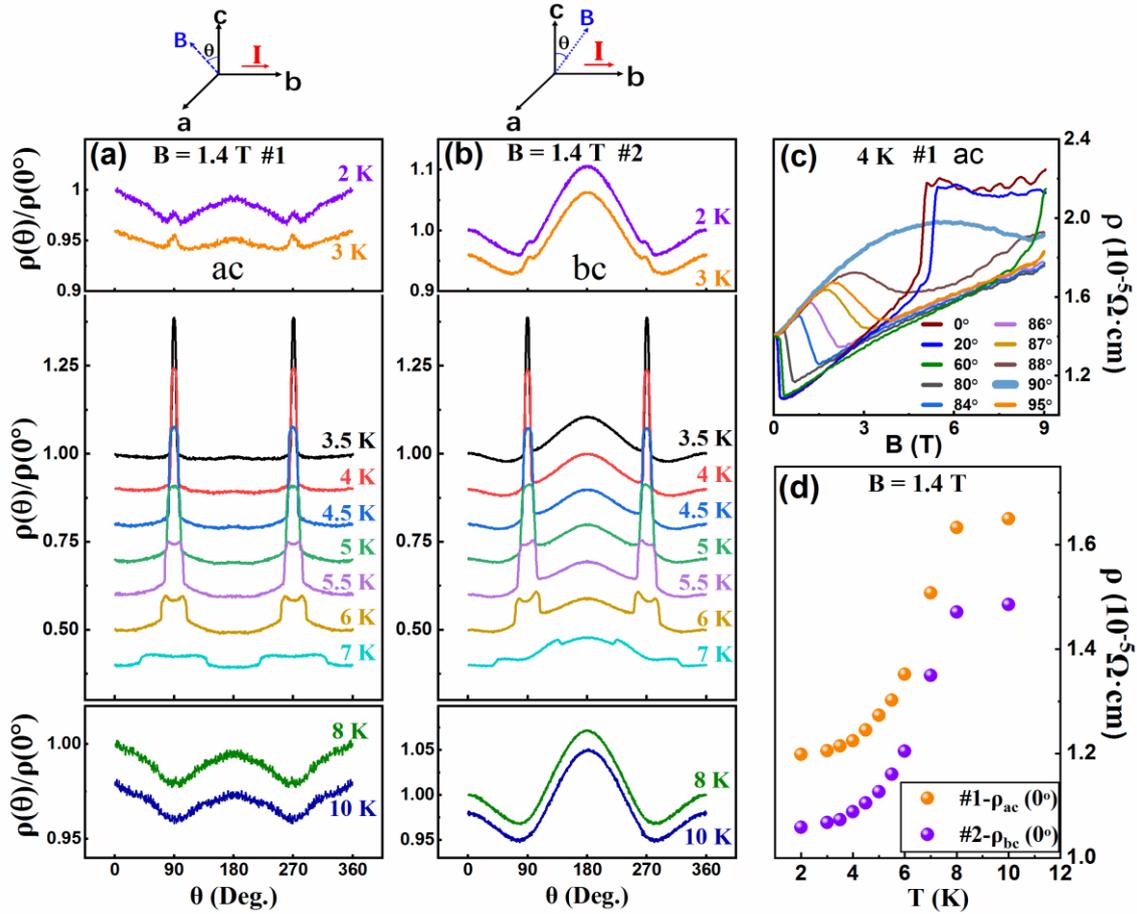

**FIG. 5.** Angular magnetoresistivity of NdAlSi and its evolution with temperature. (a) and (b) Magnetic field tiling angle dependent resistivity ρ (θ) normalized by corresponding ρ (0°) values, with current perpendicular and within the magnetic field rotation plane, respectively. The measurement configurations are shown on top. Data are grouped into sub-panels by temperature ranges according to magnetic phases at zero field. Curves are vertically shifted for clarity. (c) Magnetoresistivity ρ (B) at 4 K for different magnetic field orientation in ac plane for sample #1. (d) Temperature dependent ρ (0°) extracted from ρ (θ) at 1.4 T for sample #1 and #2, which were cut from the same piece of single crystal.

The effects on transport properties of NdAlSi by intrinsic magnetic moments have a novel manifestation on the angular magnetoresistance (AMR). As shown in Fig. 5(a), with magnetic field stabilized at 1.4 T, the sample was rotated in a plane containing the magnetic field direction and perpendicular to the current direction. Starting from 10 K, firstly the isothermal AMR curves are varying smoothly, then a bump centering at B⊥c develops, it becomes narrower and taller with cooling, reaches maximum as a narrow peak at 3.5 K, and then drops into a tiny peak at lower temperature. A similar evolution is also observed as the field rotated in the plane defined by directions of B and I, as shown in Fig. 5(b). A sharp peak in AMR can be ascribed to either the Fermi-surface topological effect in quasi-two-dimensional conductor [47-48], or the existence of high resistance domain walls resulting from the broken magnetic symmetry and Fermi surface mismatch in a nearly nodal electronic structure, as reported in the magnetic nodal semimetal $CeAlGe_{1-x}Si_x$ [26]. Obviously, such a strikingly enhanced $\rho(\theta)$ in NdAlSi confined with the magnetic field along a high symmetry axis of the NdAlSi crystal is reminiscent of that observed in $CeAlGe_{1-x}Si_x$ [26], in which a sharp decrease in AMR was realized when the constant external magnetic field is rotated away from the easy magnetization a-axis. The phase diagram of CeAlGe for B//a-axis has a region with canted magnetic moments that organized into domains. In the aforementioned reference, it is suggested that the real space domain walls are of high resistance since in momentum space the Fermi surfaces of different domains have mismatch, while a semimetal with strong magnetic anisotropy, a small Fermi surface close to node, and strong spin-orbit coupling all helps. How stiff the domain walls could hold against rotation of magnetic field determines the width of the peak in AMR. Here, the peaked resistivity in NdAlSi occurs at the magnetic field direction perpendicular to the easy magnetization c-axis, which is in marked contrast to the configuration of CeAlGe. However, we could still make an explanation in a similar way. Without considering Landau quantization effects on AMR, and starting from the **B**//c phase diagram we established in Fig. 3. For B = 1.4 T, the compound is within the FIM region, as evidenced by the extracted temperature-dependent magnetoresistance in Fig. 5(d). Since the magnetization for $\mathbf{B}\perp c$ is almost linear, a deduction is for that direction there would not exist a phase boundary in the low field region. By rotating the sample away, the projected B along $\mathbf{B}//c$ becomes smaller and is zero at $\mathbf{B}\perp c$, thus effectively the compound moves from the FIM region crossing the phase boundary line into the AFM region. Fig. 5(c) demonstrates the fast change of phase boundary when the magnetic field is close to $\mathbf{B}\perp c$. The narrowing of the bump with cooling follows the phase boundary line, and the increase in amplitude is in accordance with the isothermal MR for B//c as in Fig. 2(c). It is now clear that such unusual AMR is closely linked to the temperature-magnetic field phase diagram. The apparent difference between CeAlGe and NdAlSi regarding whether the peak in AMR is at the easy magnetization direction originates from the difference in the configuration of magnetic moments in these two compounds [26, 39]. For CeAlGe, with B//a the canted phase lies in the intermediate field range between the low field ferrimagnetic and high field ferromagnetic regions [26], while for NdAlSi at an intermediate temperature the low field AFM region has larger in plane ($\mathbf{M}\perp c$) moment components than the adjacent high field FIM region [39]. In addition to the high resistance domain wall, several other sources, including Fermi surface reconstruction happen as the phase boundary crossing taking place, the opening of a gap on Fermi surface leading to reduced carrier density, and the releasing of spin-flip scattering channels in the AFM region, could also contribute to the enhanced resistance. But after all, the tunability of magnetic structure of semimetallic NdAlSi and related compounds is the key in realizing the singular AMR behavior [39]. In addition, in the configuration that sample rotates the plane defined by B and I, an asymmetry between 0º and 180º AMR exists over the measured temperature range. It is reproduced in several pieces of crystals and not likely due to measurement and/or data process errors, but possibly due to a spin valve mechanism [49], which has also been observed in a bulk material $SrCo_6O_{11}$ [50]. For NdAlSi, rotating an external magnetic field of 1.4 T would drive the f-electron moments to re-arrange, thus changes the relative angle between adjacent moments and affects spin-flip scattering, indicating that the moments of f electrons affect the transport properties in more than one channel and even into the paramagnetic region.

## IV. CONCLUSION

The magnetization, magneto-transport and electronic band structure of MWSM candidate NdAlSi are systematically investigated. A magnetic phase diagram for **B**//c is constructed, including paramagnetic, antiferromagnetic, ferrimagnetic, and field-induced polarized paramagnetic regions. An anomalous AMR appears close to the border between antiferromagnetic and ferrimagnetic regions, which shows up as a sharp peak in a narrow angle range instead of a gradual change as observed in the conventional AMR. Our findings imply that the tunable topological and magnetic properties in 4f-electron based magnetic topological systems are uniquely suited to manipulating the topological states and realizing novel quantum phenomena, and call for further in-depth investigations.

## V. AUTHOR CONTRIBUTION AND ACKNOWLEDGMENTS

J. F. Wang, Q. X. Dong and Z. P. Guo contributed equally to this work. The authors acknowledge Z. Y. Mi, Y. Y. Wang, H. C. Zhao and W. L. Zhu for valuable discussions. This work is supported by the National Natural Science Foundation of China (Grant No. 11874417, U1504107), the Strategic Priority Research Program (B) of Chinese Academy of Sciences (Grant No. XDB33010100, XDB33020100).

## VI. APPENDIX

Figures 6-12 and tables I-II present additional data, including the crystal structure refinement, the Curie-Weiss fitting, magnetic field dependent M(T), SdH analysis at different magnetic phases regions, and the band structures of NdAlSi in the PPM and FIM magnetic states without spin-orbit coupling (SOC), and PM states with SOC.

TABLE I. Refinement of the crystal structure and atomic information of **NdAlSi single crystal** at room temperature.

| Chemical formula: NdAlSi |
|---|
| Crystal structure: Tetragonal LaPtSi type |
| Space group: $I4_1md$ (No. 109) |
| Lattices parameters: **a = b = 4.201(8) Å, c = 14.516(2) Å**, V = 256.28(5) Å$^3$, α = β = γ = 90° |
| Z: 4 |
| Final R indices (all data): $R_p$ = 2.3%, $R_w$ = 5.8% |
| Goodness-of-fit on F$^2$: 1.226 |

| Atom | Wyck. | x/a | y/b | z/c | $U_{11}$ | $U_{22}$ | $U_{33}$ | $U_{23}/U_{13}/U_{12}$ |
|---|---|---|---|---|---|---|---|---|
| Nd | 4a | 0.5 | 0.5 | 0.503 | 0.0108 | 0.0105 | 0.015 | 0/0/0 |
| Al | 4a | 0.5 | 0 | 0.3379 | 0.017 | 0.047 | 0.01 | 0/0/0 |
| Si | 4a | 0 | 0 | 0.4214 | 0.005 | 0.005 | 0.027 | 0/0/0 |

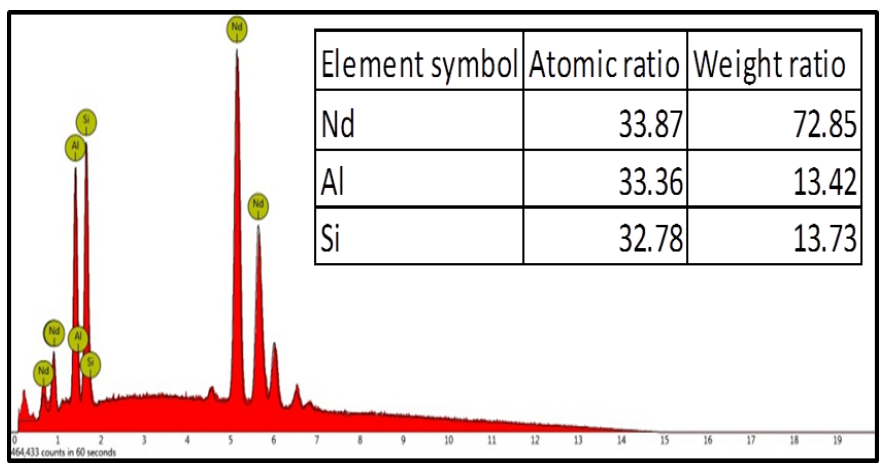

**FIG. 6.** A typical elements analysis result determined by EDX of NdAlSi single crystal.

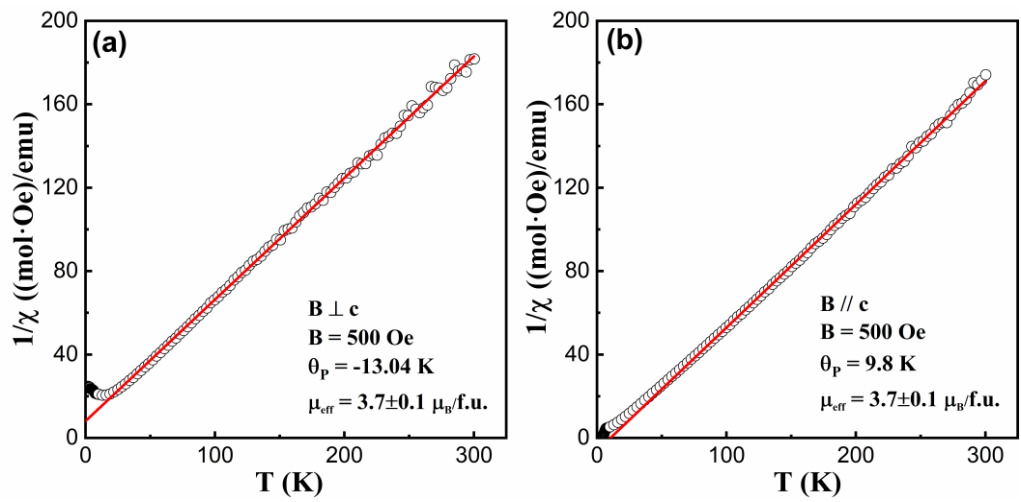

**FIG. 7.** The Curie-Weiss fitting of $\chi^{-1}(T)$ for (a) $B \perp c$ and (b) $B // c$. The fitting range is from 50 K to 300 K.

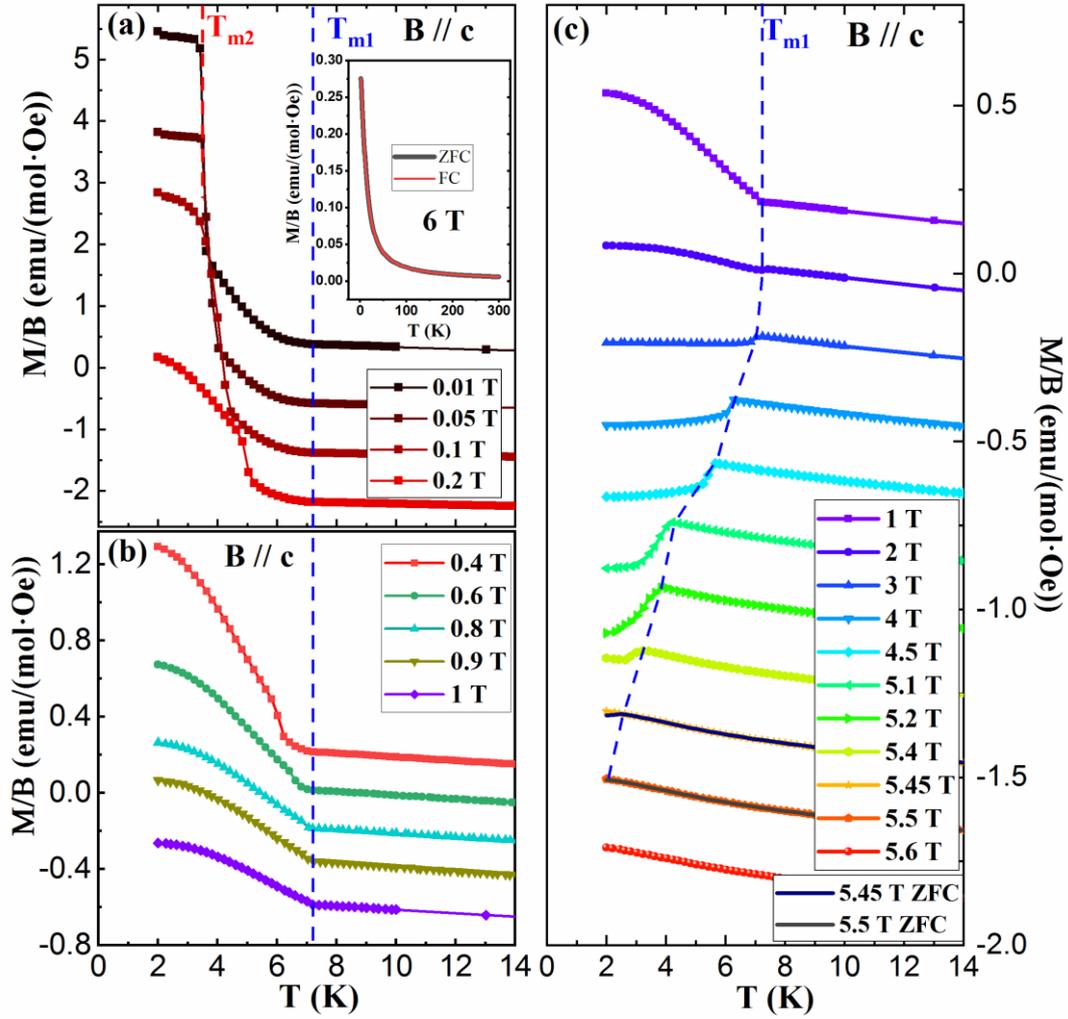

**FIG. 8.** Evolution of M (T) for B//c at magnetic fields from (a) 0.01-0.2 T, (b) 0.4-1 T and (c) 1-5.6 T. For clarity, the curves of Figs. 6(a)-(c) are shifted downward by 0.8, 0.2 and 0.2 emu/(mol·Oe), respectively. The inset in panel a is the ZFC/FC M(T) data taken at 6 T. The most significant feature in χ (T) is that a sign change of the magnetization jump (upturn of FC changes to downturn) occurring right at a critical field where the ferrimagnetism was found to disappear

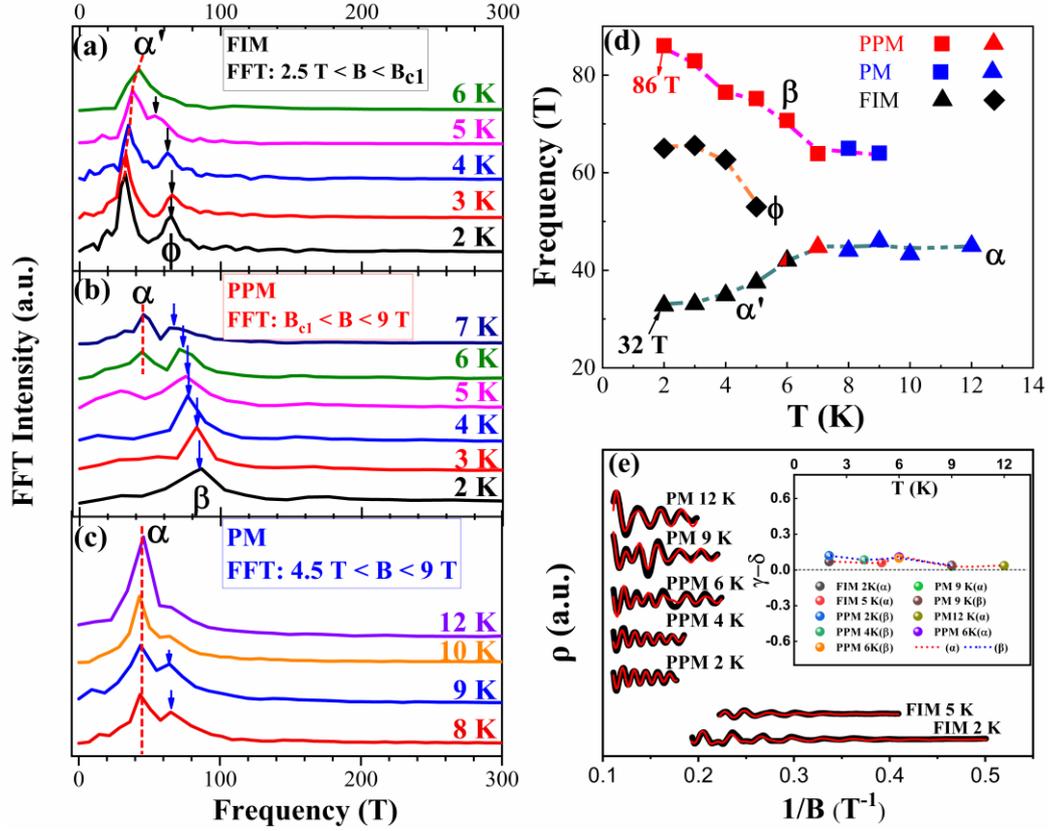

**FIG. 9.** SdH analysis for different magnetic phase regions for B//c. (a) The FFT spectra of the oscillating magnetoresistivity (MR$_{OSC}$) in the FIM phase. The FFT spectra of MR$_{OSC}$ in the PPM (b) and the PM (c) phases, the spectra are vertically shifted for clarity. (d) Evolution of extracted oscillation frequencies with temperature. (e) Fitting of MR$_{OSC}$ by the standard Lifshitz-Kosevich (LK) formula $MR_{osc} \sim \sum_{i=\alpha,\beta} A_i(B,T)\cos[2\pi(\frac{F_i}{B} + \gamma_i - \delta_i)]$. Inset: The extracted phase $|\gamma - \delta|$ as a function of temperature for three magnetic phases. A non-zero geometric phase is resultant.

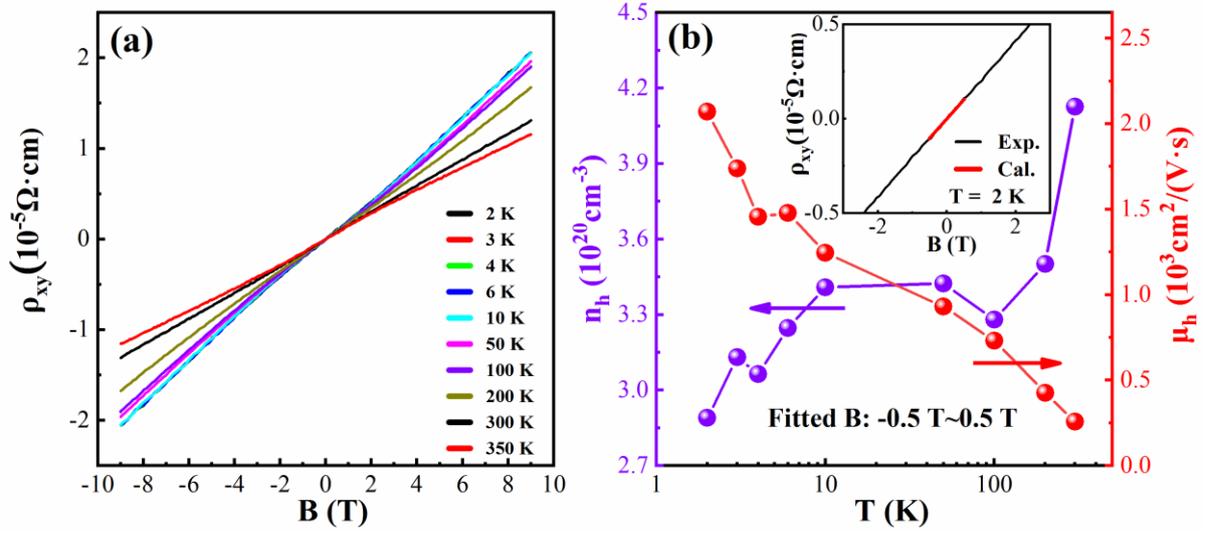

**FIG. 10.** (a) Isothermal Magnetic Field dependent Hall resistivity. (b) Temperature dependence of the Carrier concentration and Hall mobility extracted by the single band model from the low field region around zero.

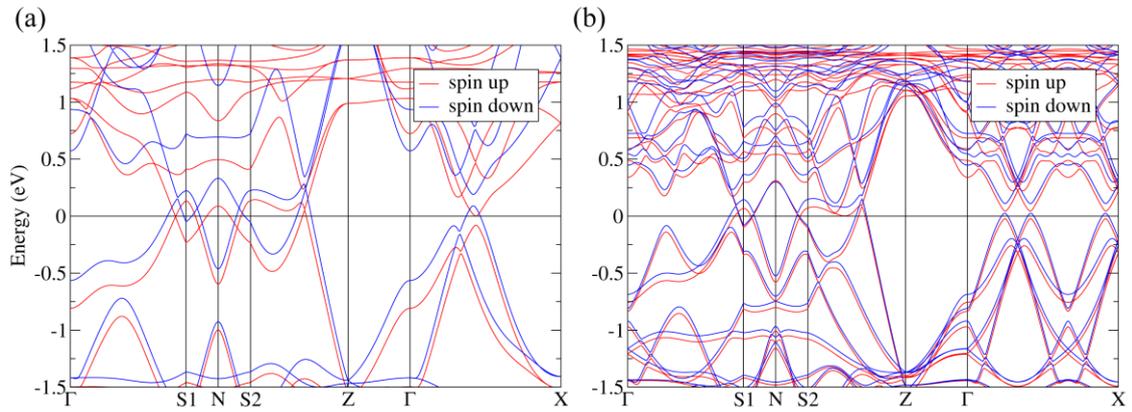

**FIG. 11.** The calculated band structures of NdAlSi in (a) the PPM state and (b) the FIM state without spin-orbit coupling (SOC), respectively. The spin-up and spin-down bands are marked by different colors.

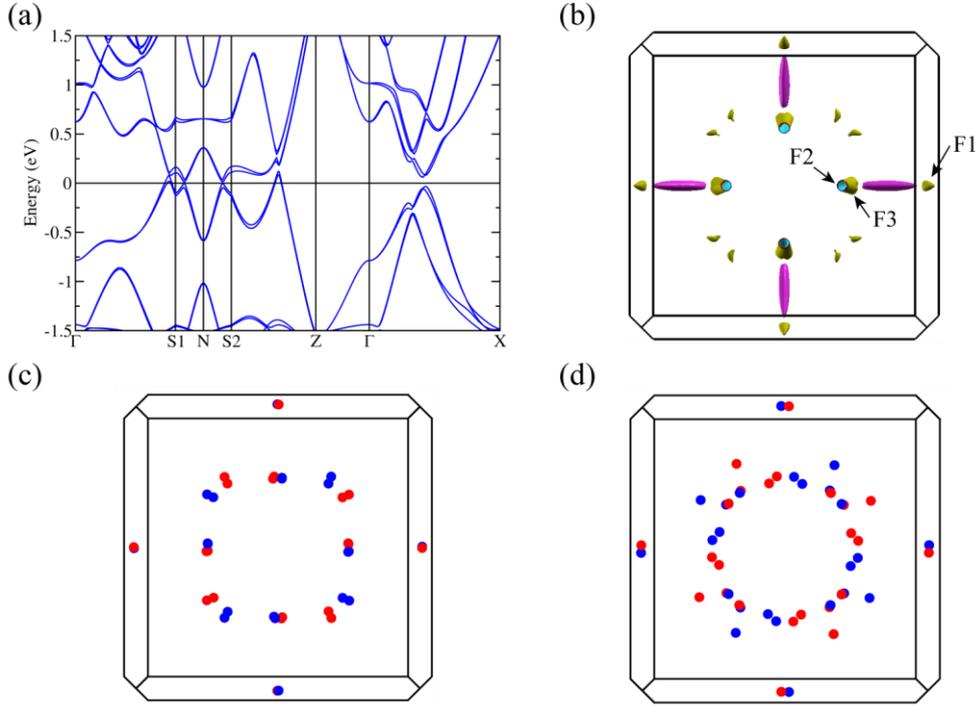

**FIG. 12.** (a) The calculated band structure of NdAlSi in the PM state including SOC, and (b) The corresponding Fermi surfaces. The locations of extreme orbits are marked by F1, F2 and F3. (c) The Weyl points in the PM state, red and blue dots mark chirality of +1 and -1, respectively. (d) The Weyl points in PPM state with the same notion.

**TABLE** II. The calculated quantum oscillation frequencies and location of extreme cross sections in the PPM and the PM states.

|  | PPM | | PPM ($E_F$ +20 meV) | | PM | |
|---|---|---|---|---|---|---|
|  | Frequency (T) | Location | Frequency (T) | Location | Frequency (T) | Location |
| F1 | 56 | $k_z = 0$ | 35 | $k_z = 0$ | 24 | $k_z = 0$ |
| F2 | 92 | $k_z = \pi$ | 76 | $k_z = \pi$ | 27 | $k_z = \pi$ |
| F3 | 134 | $k_z = 0.6\pi$ | 93 | $k_z = 0.62\pi$ | 40 | $k_z = 0.7\pi$ |